\documentclass[12pt]{article}
\usepackage{amsmath,amsfonts, amssymb}
\usepackage{graphicx}
\usepackage{longtable}
\usepackage{url}
\usepackage{tocloft}
\makeatletter
 
  \@addtoreset{equation}{section}
\makeatother
\def\beqar {\begin{eqnarray}}
\def\eeqar {\end{eqnarray}}
\def\beq {\begin{equation}}
\def\eeq {\end{equation}}
\def\A{{\cal A}}

\def\D{{\cal D}}

\def\al{\alpha}
\def\bt{\beta}
\def\del{\delta}

\def\ga{\gamma}
\def\Ga{\Gamma}

\def\La{\Lambda}
\def\om{\omega}
\def\Om{\Omega}
\def\th{\theta}

\def\d{\partial}
\def\bd{{\bar \partial}}

\def\ba{{\bar a}}
\def\bz{{\bar z}}

\def\bom{{\bar \omega}}

\def\hf{\frac{1}{2}}

\def\<{\langle}
\def\bra{\langle}
\def\>{\rangle}
\def\ket{\rangle}

\def\re{{\rm Re}}
\def\im{{\rm Im}}

\def\C{\mathbb{C}}

\def\Z{\mathbb{Z}}

\def\H{\mathbb{H}}

\begin{document}

\begin{titlepage}
\null\vspace{-62pt} \pagestyle{empty}
\begin{center}
\vspace{0.8truein}

{\Large\bf
Some properties of zero-mode wave functions
\\
\vspace{.35cm}
in abelian Chern-Simons theory on the torus
} \\

\vspace{1.0in} {\sc Yasuhiro Abe} \\
\vskip .12in {\it Cereja Technology Co., Ltd.\\
3-11-15 UEDA-Bldg. 4F, Iidabashi   \\
Chiyoda-ku, Tokyo 102-0072, Japan }\\
\vskip .07in {\tt abe@cereja.co.jp}\\
\vspace{1.3in}

\centerline{\large\bf Abstract}
\end{center}

\noindent
In geometric quantization a zero-mode wave function
in abelian Chern-Simons theory on the torus
can be defined as $\Psi [ a, \ba ] = e^{- \frac{K(a, \ba)}{2}} f (a)$
where $K(a ,\ba )$ denotes a K\"ahler potential for the zero-mode
variable $a \in \C$ on the torus.
We first review that the holomorphic wave function $f(a)$  can be described in terms of
the Jacobi theta functions by imposing gauge invariance on $\Psi [ a, \ba ]$
where gauge transformations are induced by doubly periodic translations of $a$.
We discuss that $f(a)$ is quantum theoretically characterized by
({\it i}) an operative relation in the $a$-space representation
and ({\it ii}) an inner product of $\Psi [ a, \ba ]$'s including
ambiguities in the choice of $K(a ,\ba )$.
We then carry out a similar analysis on the gauge invariance
of $\Psi [ a , \ba ]$ where the gauge transformations are
induced by modular transformations of the zero-mode variable.
We observe that$f(a)$ behaves as a modular form of weight 2 under the condition of $|a|^2 = 1$,
{\it i.e.}, $\left. f \left( - \frac{1}{a} \right) =  a^2 f(a) \right|_{|a|^2 = 1}$.
Utilizing specific forms of $f(a)$
in terms of the Jacobi theta functions, we further investigate
how exactly $f(a)$ can or cannot be interpreted as the modular form of weight 2;
we extract conditions that make such an interpretation possible.

\end{titlepage}
\pagestyle{plain} \setcounter{page}{2} 

\tableofcontents
\vskip 1.2cm

\section{Introduction}

It has been known for a long time that
the holomorphic zero-mode wave functions in abelian Chern-Simons (CS) theory on the torus can
be expressed in terms of a Jacobi theta function
in geometric quantization \cite{Bos:1989wa,NairBook,Nair:2016ufy};
similar results were also obtained
in the cases of the non-abelian CS theory \cite{Bos:1989kn,Labastida:1990bt}.
For a standard mathematical literature on
the geometric quantization of CS theory, see \cite{ADW:1991}.

As explained long ago in the mathematical literature \cite{Mumford:tata},
these theta functions can be described as modular forms.
Also, given the everlasting importance of the modular invariance
in developments of string theory and two-dimensional
conformal field theory, it would be
desirable to deepen the understanding of the above mentioned result
on the zero-mode wave function particularly in relation to
modular transformations.
One of the main motivations for this paper is to carry out such a study.

The organization of the paper is as follows.
In the next section, following \cite{Bos:1989wa,NairBook,Nair:2016ufy},
we briefly review geometric quantization of abelian CS theory on the torus.
We indicate that crucial ingredients of the quantization
(such as a K\"ahler potential,
a symplectic potential, and a wave function)
are all derived from a K\"ahler form for the zero-mode variables
defined on the torus of interest.
We emphasize that there exist ambiguities in the choice of the K\"ahler potential.
At a prequantum level the zero-mode wave function can be interpreted as
a wave function of a complex scalar field that couples
to the symplectic potential.
The ordinary quantum wave function is obtained by imposing a polarization
condition on the prequantum wave function.
We also present an explicit form of an inner product for
the (quantum) zero-mode wave functions.

In section 3 we consider gauge transformations of the
symplectic potential induced by the doubly periodic translations
of the zero-mode variable.
We then impose gauge invariance on the zero-mode wave function
and see its consequences.
We find that holomorphic part of the zero-mode wave function can be described
in terms of a Jacobi theta function, reproducing the results in
\cite{Bos:1989wa,NairBook,Nair:2016ufy}. We also find an alternative
representation of the holomorphic zero-mode wave function in terms of
another version of the Jacobi theta functions.
Along the way, we discuss that the holomorphic zero-mode wave functions
are characterized by an operative relation
(in the holomorphic-coordinate-space representation) and the inner product
of the zero-mode wave functions (which includes ambiguities in the choice
of the K\"ahler potential).

In section 4 we apply the same analysis, {\it i.e.},
imposition of the gauge invariance on the zero-mode wave function
where the gauge transformations are induced by the modular transformations
of the zero-mode variable.
We consider the modular transformations of the zero-mode
variable, rather than those of the modular parameter of the torus.
The relevant modular transformations are thus
different from the conventional ones used in string theory and
conformal field theory.
Since the modular $T$-transformation is included in the
doubly periodic translations, we focus on the modular $S$-transformation.
Note that the modular $S$-transformation is not globally well-defined
as a map from the torus to itself since the transformation
does not preserve the periodicity of the zero-mode variable.
This fact may cause a crucial defect in the present analysis but
what we consider is the gauge invariance of the wave function
where the gauge transformations (of the symplectic potential) is
induced by the modular $S$-transformation of the zero-mode variable;
for concrete expressions, see (\ref{4-6})-(\ref{4-8}).
The transformations of interest are thus considered
local in terms of the zero-mode variable
and the present analysis makes sense at local level.
We observe that under certain conditions
the holomorphic wave function behaves as a 
quantum version of a modular form of weight 2.
(Some basic facts on modular forms are reviewed in the appendix, following mathematical
textbooks \cite{Koblitz:1993bk,Kurokawa:2005bk,Lozano-Robledo:2011bk}.)

In section 5 we further investigate this relation, 
utilizing specific forms of the holomorphic wave function in
terms of the Jacobi theta functions. 
We study how exactly the holomorphic wave function can or cannot
be interpreted as the modular form of weight 2; we clarify
conditions on $f(a)$ and $a$ that make such an interpretation possible.
Lastly, in section 6 we present brief conclusions.

\section{Zero-mode wave functions in abelian CS theory on the torus}

We first briefly review how to construct zero-mode wave functions of
abelian Chern-Simons (CS) theory on the torus in the context of geometric quantization,
following \cite{Bos:1989wa,NairBook,Nair:2016ufy}.

\vskip 0.5cm \noindent
\underline{Basics on the torus parametrization}

The torus can be described in terms of two real coordinates $\xi_1$, $\xi_2$,
satisfying the periodicity condition $\xi_r \rightarrow \xi_r + {\rm (integer)}$
where $r=1,2$.
In other words, $\xi_r$ take real values in $0 \le \xi_r \le 1$,
with the boundary values $0$, $1$ being identical.
Complex coordinates of the torus can be
parametrized as $z = \xi_1 + \tau \xi_2$ where $\tau \in \C$ is
the modular parameter of the torus.
By definition, we can impose the doubly periodic condition on $z$.
Namely, functions of $z$ are invariant under the doubly periodic translations
\beq
    z \, \rightarrow \, z + m + n \tau
    \label{2-1}
\eeq
where $m$ and $n$ are integers.
Notice that we can absorb the real part of $\tau$ into $\xi_1$
without losing generality.
In the following, we then assume $\re \tau = 0$, {\it i.e.},
\beq
    \tau \, = \, \re \tau + i \im \tau \, =  \, i \im \tau  \, := \, i \tau_2
    \label{2-2}
\eeq
with $\tau_2 > 0$.

The torus has a holomorphic one-form $\om = \om (z) dz $, satisfying
\beq
    \int_\al \om = 1 \, , ~~~ \int_\bt \om = \tau = i \tau_2
    \label{2-3}
\eeq
where the integrations are made along two non-contractible cycles
on the tours, which are conventionally labeled as $\al$ and $\bt$ cycles.
The one-form $\om$ is a zero mode of the anti-holomorphic derivative
$\d_\bz = \frac{\d}{\d \bz}$. We can assume $\om (z) = 1$.
In terms of $\om$ a gauge potential of abelian CS theory on the torus
can be parametrized as
\beq
    A_\bz  =   \d_\bz \th +  \frac{ \pi \bom }{ \tau_2 } a
    \label{2-4}
\eeq
where $\th$ is a complex function $\th (z , \bz )$ and
$a$ is a complex number corresponding to the value of
$A_\bz$ along the zero mode of $\d_z$.
The abelian gauge transformations can be represented by
\beq
    \th \, \rightarrow \, \th + \chi
    \label{2-5}
\eeq
where $\chi$ is a complex constant or a phase factor of the $U(1)$ theory.
With a suitable choice of $\chi$ we can parametrize the
gauge potential solely by the zero-mode contributions, $a$ and its complex
conjugate $\ba$:
\beq
    A_{z} \, = \,   \frac{\pi \om}{ \tau_2 } \ba \, , ~~~~
    A_{\bz} \, = \,  \frac{\pi \bom}{\tau_2 } a \, .
    \label{2-6}
\eeq

Since the complex variable $a$ is defined on the torus it is natural to
require that physical observables of the zero modes are also invariant under
the doubly periodic translations
\beq
    a \, \rightarrow \, a + m + i n \tau_2
    \label{2-7}
\eeq
where $m$ and $n$ correspond to the
winding numbers along the $\al$ and $\bt$ cycles, respectively.
The physical configuration space of the zero mode is given by
\beq
    \cal{C} \, = \, \frac{\C}{ \Z + {\it i}{\rm \tau_2}  \Z } \, .
    \label{2-8}
\eeq
This is nothing but a complex torus, with the modular parameter being $i \tau_2$.
Mathematically, this space is known as an {\it abelian variety} over
the field of complex numbers.

\vskip 0.5cm \noindent
\underline{Key ingredients in geometric quantization}

Geometric quantization provides a powerful quantization scheme
when a phase space of a physical system
is given by a K\"ahler manifold \cite{NairBook,Nair:2016ufy}.
The K\"ahler manifold has both symplectic and complex structures.
The symplectic structure takes origin from the classical
physics (to be quantized) while the complex structure makes
it automatic to realize irreducibility (or polarization) of operators.
The torus $S^1 \times S^1$ is a K\"ahler manifold with
the simplest nontrivial topology. A physical system on the torus
can be described in terms of an inherent zero-mode variable.
As discussed above, this is particularly true in the case of abelian CS
theory on the torus where the contributions from ordinary
non-zero-mode part of the abelian gauge potential can be gauged away.

The zero-mode dynamics can then be encoded by a K\"ahler form
for the zero-mode part of the CS gauge potentials (\ref{2-6}), {\it i.e.},
\beq
    \Om^{(\tau_2 )} \, = \, \frac{l}{2 \pi} da \wedge d\ba \int_{z,\bz}
    \frac{ \pi \bom}{\tau_2}    \wedge  \frac{ \pi \om}{\tau_2}
    \, = \, i \frac{\pi l}{ \tau_2} da \wedge d \ba
    \label{2-10}
\eeq
where the integral is taken over $dz d \bz$ and
$l$ is the level number associated to the abelian CS theory.
We here use the normalization of $\om$ and $\bom$ given by
\beq
    \int_{z,\bz} \bom \wedge \om  \, = \, i 2  \tau_2 \, .
    \label{2-11}
\eeq
A K\"ahler potential $K( a, \ba)$ associated with the zero-mode K\"ahler form
$\Om^{(\tau_2 )}$ is defined as
\beq
    \Om^{(\tau_2 )} \, =  \,   i \d \bd K (a , \ba )
    \label{2-12}
\eeq
where $\d$, $\bd$ denote the Dolbeault operators.
This definition leads to
\beq
    K (a, \ba ) \, = \, \frac{\pi l}{ \tau_2 } a \ba + u (a ) + v ( \ba )
    \label{2-13}
\eeq
where $u(a)$ and $v( \ba )$ are purely holomorphic and
anti-holomorphic functions, respectively. These functions
represent {\it ambiguities} in the choice of $K (a, \ba )$.

A symplectic potential (or a canonical one-form) $\A^{(\tau_2 )}$
corresponding to the K\"ahler form $\Om^{(\tau_2 )}$ is defined as
\beq
    \Om^{(\tau_2 )} \, =  \, d \A^{(\tau_2 )} \, .
    \label{2-14}
\eeq
Under a canonical transformation the K\"ahler form $\Om^{(\tau_2 )}$
does not change but the symplectic potential transforms as
\beq
     \A^{(\tau_2 )} \, \rightarrow \,  \A^{(\tau_2 )} + d \La
     \label{2-15}
\eeq
where $\La$ is a function of $( a , \ba )$.
In other words, the symplectic potential $\A^{(\tau_2 )}$
undergoes a $U(1)$ gauge transformation.

In the program of geometric quantization a quantum wave function
arises from a {\it prequantum}
wave function $\Psi [ A_\bz ]$ which is a function of $( a , \ba )$
in the present case.
Mathematically we can state that the prequantum wave function is a section
of line bundle on the torus with curvature $\Om^{(\tau_2 )}$.
This means that $\Psi [ A_\bz ]$ transform as
\beq
    \Psi [ A_\bz ] \, \rightarrow \,
    \Psi^\prime [ A_\bz ] \, = \, e^{i \La} \Psi [ A_\bz ]
    \label{2-16}
\eeq
under the $U(1)$ gauge transformation (\ref{2-15}).
At the prequanum level $\Psi [ A_\bz ]$ can be interpreted as a wave
function of a complex scalar field that couples
to the symplectic potential $\A^{(\tau_2 )}$.

In order to obtain a {\it quantum} wave function we need to impose
the  so-called polarization condition on $\Psi [ A_\bz ]$, {\it i.e.},
\beq
    \left( \d_\ba + \frac{1}{2} \d_\ba K \right) \, \Psi [ A_\bz ] \, = \, 0
    \label{2-17}
\eeq
where $K = K ( a , \ba )$ is the zero-mode K\"{a}hler potential in (\ref{2-13}).
The polarization condition leads to the specific form
\beq
    \Psi[ A_\bz ] \, = \, e^{-\frac{K}{2}} \psi[ A_\bz ]
    \label{2-18}
\eeq
where $\psi [ A_\bz ]$ is a holomorphic function of $A_\bz$.
In the present case the physical variables are given by $(a , \ba )$
so that the wave function can be expressed as
\beq
    \Psi[ A_\bz] \, := \, \Psi[a , \ba ] \, = \, e^{-\frac{K (a,\ba)}{2}} f(a)
    \label{2-19}
\eeq
where $f(a)$ is a function of $a$. We call $f(a)$ a {\it holomorphic zero-mode
wave function}.
Notice that we here define $\Psi[ a , \ba ]$ with $K ( a , \ba )$ including
the above-mentioned ambiguities in its choice.

Let us fix the K\"ahler potential by $K_0 := \frac{\pi l}{\tau_2 } a \ba$.
The symplectic potential can also be chosen as
\beqar
    \A^{( \tau_2 )} & = & \frac{l}{4 \pi}
    \int_{z , \bz} \left(
    \frac{ \pi \bom a}{\tau_2} \wedge \frac{ \pi \om}{ \tau_2} d \ba
    - \frac{ \pi \om \ba }{\tau_2}  \wedge \frac{ \pi \bom  }{ \tau_2} d a
    \right)
    \nonumber \\
    & = & i \frac{ \pi l}{2 \tau_2 } ( a d \ba + \ba d a)
    \nonumber \\
    & := &  \A^{( \tau_2 )}_{\ba} d \ba +  \A^{( \tau_2 )}_{a} da  \, .
    \label{2-20}
\eeqar
In these choices the polarization condition (\ref{2-17}) can be expressed
as $\D_\ba \Psi = 0$ where $\D_\ba := \d_\ba + \frac{1}{2} \d_\ba K_0
= \d_\ba - i \A^{( \tau_2 )}_{\ba}$ is given in a form of a covariant derivative.
As mentioned before, under a canonical transformation
$\A^{( \tau_2 )}$ transforms as $\A^{(\tau_2 )} \rightarrow  \A^{(\tau_2 )} + d \La $
while $\Om^{( \tau_2 )}$ remains the same. Thus, from the definition (\ref{2-12}),
a change of $K ( a , \ba )$ under the canonical transformation
should be absorbed in the terms of $u(a) + v (\ba )$ in (\ref{2-13}).
In other words, the form of K\"ahler potential dose not change under
the $U(1)$ gauge transformations of $\A^{(\tau_2 )}$
Thus, it is appropriate to define the polarization condition
in terms of $K ( a , \ba )$ as in (\ref{2-17}) rather than
in terms of $\A^{( \tau_2 )}_{\ba}$.

This allows us to choose $\A^{( \tau_2 )}$ independently of
the polarization condition. For example, we can define $\A^{( \tau_2 )}$  as
\beq
    \A^{( \tau_2 )}  \, = \, i \frac{\pi l}{ \tau_2} \ba d a \, .
    \label{2-21}
\eeq
This is a suitable choice in the $a$-representation space.
In fact, we can regard this as an irreducible representation of $\A^{( \tau_2 )}$
in the $a$-space since, as a general feature in quantum theories,
irreducibility of an operator can be realized by a (holomorphic) polarization
condition.


An inner product of the zero-mode wave functions $\Psi[a , \ba ]$ in (\ref{2-19})
can be expressed as
\beqar
    \< \Psi | \Psi^\prime \> & = &
    \int d\mu(a,\ba)  \, \overline{\Psi[ A_\bz ]} \Psi^\prime [ A_\bz ]
    \nonumber \\
    & = &
    \int d\mu(a,\ba) \, e^{-K(a,\ba)} \, \overline{f(a)} f^\prime (a)
    \label{2-22}
\eeqar
where $\overline{f(a)}$ is the complex conjugate of  $f (a)$.
The integral is taken over the complex plane $\C$ although, strictly speaking,
the zero-mode variable is defined on ${\cal C} = \frac{\C}{ \Z + {\it i}{\rm \tau_2}  \Z }$
as discussed in (\ref{2-8}).
In the above expression we then need to
impose the doubly periodic condition on $a$ by hand. In this sense,
the integral over $d \mu (a , \ba )=  da d \ba $ is the same as the
integral over $dz d \bz$ in (\ref{2-10}, \ref{2-11}).
Lastly, we notice that, as a consequence of
the geometric quantization \cite{NairBook,Nair:2016ufy},
an action of the derivative $\frac{\d}{\d a}$ on $f(a)$ leads to
the factor of $\frac{\pi l}{\tau_2 } \ba$.

\section{Gauge transformations induced by doubly periodic translations}

As mentioned in the introduction, any complex functions on the torus
may obey the double periodicity condition. Thus, at classical level,
we can naturally expect that the holomorphic zero-mode
wave function $f(a)$ is invariant under the doubly periodic translations
\beq
    a \, \rightarrow \, a  + m + i n \tau_2 \, .
    \label{3-1}
\eeq
Quantum theoretically, however, this invariance is not
necessarily guaranteed. In this section we consider
this problem by use of the quantum wave function $\Psi[a , \ba ]$ in (\ref{2-19}).
Our strategy is to impose a gauge invariance
of the zero-mode wave function $\Psi[a , \ba ] = \Psi^\prime [a , \ba ]
= e^{i \La}\Psi[a , \ba ] $ under the gauge transformation
$\A^{(\tau_2 )} \rightarrow \A^{(\tau_2 )} + d \La$ which is induced by
the doubly periodic translation of $a$.

\vskip 0.5cm \noindent
\underline{Connection to the Jacobi theta functions}

Inspired by the results in \cite{Bos:1989wa,NairBook,Nair:2016ufy},
we now chose the symmplectic potential and the K\"ahler potential by
\beqar
    \A^{(\tau_2 ) }_{1} &=& - \frac{i \pi l}{2 \tau_2} ( \ba - a) d( \ba + a) \, ,
    \label{3-2} \\
    K_1 &=&  - \frac{\pi l}{2 \tau_2 } ( \ba - a)^2 \, = \, K_0
    - \frac{\pi l}{ 2 \tau_2} ( \ba^2 + a^2  )
    \label{3-3}
\eeqar
where, as before, $K_0 = \frac{\pi l}{\tau_2 } a \ba$.
The gauge invariance condition for the zero-mode wave function
\beq
    \Psi_1 [ a ,\ba ] \, = \, e^{- \frac{K_1}{2} } f_1 (a)
    \label{3-4}
\eeq
is then expressed as
\beq
    e^{i \La } e^{- \frac{K_1}{2} } f_1 ( a) = e^{- \frac{K_1^\prime}{2} }
    f_1 ( a + m + in \tau_2 )
    \label{3-5}
\eeq
where $\La = - \pi l n ( \ba + a )$ and $K_1^\prime = -\frac{\pi l}{2 \tau_2}
( \ba - a -  i 2 n \tau_2 )^2$.
We here label the holomorphic function by $f_1 (a)$ to
indicate that it associates with the choice of the K\"ahler potential $K_1$.
The condition (\ref{3-5}) simplifies as
\beq
    f_1 ( a + m + in \tau_2 ) \, = \, e^{ - i 2 \pi l n a+ \pi l n^2 \tau_2} f_1 (a) \, .
    \label{3-6}
\eeq
This is reminiscent of a Jacobi theta function. Indeed, for $l= 1$
we can identify $f(a)$  as one of the Jacobi theta functions:
\beq
    \vartheta_3 ( \tau, a ) \, :=  \, \sum_{n = - \infty}^{ \infty}
    q^{\frac{n^2}{2}} y^{n}
    \label{3-7}
\eeq
where $q := e^{i 2 \pi \tau}$ and $y := e^{i 2 \pi a}$.
Under the doubly periodic translations $\vartheta_3 ( \tau, a )$
transforms as
\beq
    \vartheta_3 ( \tau, a + m + n \tau ) \, = \,
    q^{-  \frac{n^2}{2}} y^{- n} \vartheta_3 ( \tau, a ) \, .
    \label{3-8}
\eeq
In the case of $\tau = i\tau_2$, this can be expressed as
\beq
    \vartheta_3 ( i\tau_2 , a + m + i n \tau_2 )
    \, = \,
    e^{ - 2 \pi i a n + \pi \tau_2 n^2 } \vartheta_3 ( i \tau_2 , a ) \, .
    \label{3-9}
\eeq
Thus $f_1 (a)$ with $l= 1$ in (\ref{3-6})
can be identified with $\vartheta_3 ( i \tau_2 , a )$.
Apart from our setting $\tau = i \tau_2$, this result agree
with the literature \cite{Bos:1989wa,NairBook,Nair:2016ufy}.

At this stage one may wonder weather the rest of the Jacobi theta functions
can also be described in the same context.
The other Jacobi theta functions $\vartheta_j ( \tau , a)$ ($j= 1,2,4$) are
defined as
\beqar
    \vartheta_1 ( \tau, a ) & :=  & \sum_{n = - \infty}^{ \infty} i (-1)^n
    q^{\hf \left(  n - \hf \right)^2} y^{n - \hf}
    \label{3-10} \\
    \vartheta_2 ( \tau, a ) & :=  & \sum_{n = - \infty}^{ \infty}
    q^{\hf \left(  n - \hf \right)^2} y^{n - \hf}
    \label{3-11} \\
    \vartheta_4 ( \tau, a ) & :=  & \sum_{n = - \infty}^{ \infty} (-1)^n
    q^{\frac{n^2}{2}} y^{n}
    \label{3-12}
\eeqar
Note that there are different conventions in the definition of
the Jacobi theta functions; we here follow those in a recent textbook \cite{Eguchi:2015bk}.
These definitions are conventional in physics.
Analogs of (\ref{3-9}) are given by
\beqar
    \vartheta_1 ( i\tau_2 , a + m + i n \tau_2 )
    & = &    (-1)^{n+ m} e^{ - 2 \pi i a n + \pi \tau_2 n^2 } \vartheta_1 ( i \tau_2 , a )
    \label{3-13} \\
    \vartheta_2 ( i\tau_2 , a + m + i n \tau_2 )
    & = &    (-1)^{m} e^{ - 2 \pi i a n + \pi \tau_2 n^2 } \vartheta_2 ( i \tau_2 , a )
    \label{3-14} \\
    \vartheta_4 ( i\tau_2 , a + m + i n \tau_2 )
    & = &    (-1)^{n} e^{ - 2 \pi i a n + \pi \tau_2 n^2 } \vartheta_4 ( i \tau_2 , a )
    \label{3-15}
\eeqar
Changing the variable $a \rightarrow a + \hf$ in (\ref{3-6}), we can easily
find
\beq
    f_1 ( a + \hf + m + in \tau_2 ) \, = \,
    (-1)^{ln} e^{ - i 2 \pi l n a + \pi l n^2 \tau_2} f_1 (a + \hf) \, .
    \label{3-16}
\eeq
Namely, we have $ f_1 ( a + \hf ) = \vartheta_4 ( i \tau_2 , a )$ for $l= 1$.
This agrees with the relation
\beq
    \vartheta_3 ( \tau , a+ \hf ) \, = \, \vartheta_4 ( \tau, a) \, .
    \label{3-17}
\eeq

Similarly, we can check other theta-function formulae
\beqar
    \vartheta_3 ( \tau , a+ \frac{\tau}{2} ) &=& q^{-\frac{1}{8}} y^{-\hf}
    \vartheta_2 ( \tau, a) \, ,
    \label{3-18} \\
    \vartheta_3 ( \tau , a+ \frac{\tau}{2} + \hf ) &=& i q^{-\frac{1}{8}} y^{-\hf}
    \vartheta_1 ( \tau, a)
    \label{3-19}
\eeqar
($q = e^{i 2 \pi \tau}$, $y = e^{i 2 \pi a}$) as follows.
Let $t_2 (a ) = \exp( -\frac{\pi l}{4} \tau_2 + i \pi l a) f_1 ( a + \frac{i \tau_2}{2} )$.
Then from (\ref{3-6}) we find $t_2 ( a + m + i\tau_2 ) = (-1)^m
\exp ( -i 2 \pi l n + \pi l n^2 \tau_2 ) t_2 (a )$.
With (\ref{3-14}) this leads to $t_2 (a ) = \vartheta_2 (a )$ for $l=1$,
which is consistent with the formula (\ref{3-18}).
The formula (\ref{3-19}) can also be checked by showing
$t_1 (a ) = \vartheta_1 (a )$ ($l = 1$) where $t_1(a)$ is defined as
$t_1 (a) = \exp( -\frac{\pi l}{4} \tau_2 + i \pi l a) f_1 ( a + \hf + \frac{i \tau_2}{2} )$.
Note that from (\ref{3-13}, \ref{3-14}) we can also find rather trivial
relations, $f_1 ( a + \frac{m}{2n} ) = \vartheta_2 ( i \tau_2 , a)$ and
$f_1 ( a + \frac{m+n}{2n} ) = \vartheta_1 ( i \tau_2 , a)$ for $l= 1$.

\vskip 0.5cm \noindent
\underline{Alternative choices of the symplectic potential and the K\"ahler potential}

One of the reasons why we can relate $f_1 (a)$
to $\vartheta_3 ( i \tau_2 , a)$ is that
the exponent in (\ref{3-6}) is independent of $m$ while we are considering
the transformation $a \rightarrow a + m + i n \tau_2$.
This happens because of our particular choices of
$\A^{( \tau_2 )}$ and $K_1$ in (\ref{3-2}, \ref{3-3}).
In terms of $( \re a , \im a )$ these can be expressed as
$\A^{( \tau_2 )} \sim ( \im a )d( \re a )$ and $K_1 \sim ( \im a )^2$, respectively.
Motivated by this thought, we now choose
\beqar
    \A^{(\tau_2 ) }_{2} &=& - \frac{i \pi l}{2 \tau_2} ( \ba + a) d( \ba - a) \, ,
    \label{3-20} \\
    K_2 &=&   \frac{\pi l}{2 \tau_2 } ( \ba + a)^2 \, = \, K_0
    + \frac{\pi l}{ 2 \tau_2} ( \ba^2 + a^2  )
    \label{3-21}
\eeqar
and consider the gauge invariance of the zero-mode wave function
\beq
    \Psi_2 [ a ,\ba ] \, = \, e^{- \frac{K_2}{2} } \overline{f_2 (a)}
    \label{3-22}
\eeq
where, for a reason to be clarified in a moment, we define the wave function
in terms of the anti-holomorphic function  $\overline{f_2 (a)} $.
The gauge invariance condition is expressed as
\beq
    e^{i \La } e^{- \frac{K_2}{2} } \overline{f_2 (a)}  = e^{- \frac{K_2^\prime}{2} }
    \overline{ f_2 ( a + m + in \tau_2 ) }
    \label{3-23}
\eeq
where $\La = - i \frac{\pi l}{\tau_2} m ( \ba - a )$ and $K_2^\prime = -\frac{\pi l}{2 \tau_2}
( \ba + a + 2 m )^2$.
The condition (\ref{3-23}) then simplifies as
\beq
    \overline{ f_2 ( a + m + in \tau_2 ) } \, = \,
    e^{ \frac{\pi l}{\tau_2} m (2 \ba + m ) } \overline{ f_2 (a) } \, .
    \label{3-24}
\eeq
The exponent does not depend on $a$ but on $\ba$; this is why we consider the anti-holomorphic
function in the definition of the wave function in (\ref{3-22}).
The exponent is also independent of $n$, contrary to the previous case in (\ref{3-6}).

Now let us introduce an anti-holomorphic function of the form
\beq
    \overline{f_3 (a) } \, := \, e^{- \frac{\pi l}{\tau_2} \ba^2 } \, \overline{ f_2 (a) } \, .
    \label{3-25}
\eeq
From (\ref{3-24}) we find
\beq
    \overline{f_3 (a + m + i n \tau_2 ) } \, = \,
    e^{i 2 \pi l n \ba + \pi l n^2 \tau_2}  \overline{ f_3 (a) } \, .
    \label{3-26}
\eeq
Namely, we have
\beq
    f_3 ( a) \, = \, e^{- \frac{\pi l}{\tau_2} a^2 }  f_2 ( a) \, = \, f_1 ( a) \, .
    \label{3-27}
\eeq
Thus $f_3 (a) $ is also equivalent to $\vartheta_3 ( i \tau_2 , a)$ for $l = 1$.

As mentioned previously, there are ambiguities in the choice
of K\"ahler potential up to addition of holomorphic and anti-holomorphic
functions. Including these ambiguities, we may define
a zero-mode wave function in a twofold way:
\beqar
    \Psi [a ,\ba ] &=& e^{-\hf ( u_1 (a) + v_1 (\ba ) )} \Psi_1 [a ,\ba]
    \label{3-28} \\
    \Psi [a ,\ba ] &=& e^{-\hf ( u_2 (a) + v_2 (\ba ) )} \overline{\Psi_2 [a ,\ba]}
    \label{3-29}
\eeqar
where $u_i (a)$ and $v_i (\ba )$ ($i = 1,2 $) are holomorphic and anti-holomorphic functions,
respectively; we denote these by $u_i$ and $\bar{v}_i$ in the following.
Using (\ref{3-28}), we can express the inner product of
the zero-mode wave function as
\beq
    \bra \Psi | \Psi^\prime \ket \, = \,
    \int d \mu ( a, \ba ) \, e^{- \left( K_0 -  \frac{\pi l}{2 \tau_2}( \ba^2 + a^2)
    + u_1 + \bar{v}_1
    \right) }
    \overline{f_{1}(a)} f_1^\prime (a)
    \label{3-30}
\eeq
where $f_1^\prime (a)$ denotes an alternative solution
of (\ref{3-6}) with possibly different choices of $( m , n)$
from those of the other solution $f_1 (a)$.
On the other hand, using (\ref{3-29}) and (\ref{3-27}), we can similarly
express the inner product as
\beqar
    \bra \Psi | \Psi^\prime \ket
    &= &
    \int d \mu ( a, \ba ) \, e^{- \left( K_0 +  \frac{\pi l}{2 \tau_2}( \ba^2 + a^2)
    + u_2 + \bar{v}_2
    \right) }
    \overline{f_{2}(a)} f_2^\prime (a)
    \nonumber \\
    &= &
    \int d \mu ( a, \ba ) \, e^{- \left( K_0 -  \frac{\pi l}{2 \tau_2}( \ba^2 + a^2)
    + u_2 + \bar{v}_2
    \right) }
    \overline{f_{1}(a)} f_1^\prime (a) \, .
    \label{3-31}
\eeqar
These two inner products differ by the holomorphic and anti-holomorphic
functions $(u_i , \bar{v}_i )$ in the exponents. In other words,
the expressions (\ref{3-30}, \ref{3-31}) give rise to a concrete
realization of the fact that there exist ambiguities in the choice of
the K\"ahler potential up to the addition of holomorphic and anti-holomorphic
functions. The relation (\ref{3-27}) can be seen as a reflection of this fact.

\vskip 0.5cm \noindent
\underline{Characterization of the holomorphic zero-mode wave function}

We have shown that $f_1(a)$ and $f_2(a)$
are described by the Jacobi theta functions.
Under the doubly periodic translations these transform as
$f_1 ( a + m + in \tau_2 ) = e^{ - i 2 \pi l n a+ \pi l n^2 \tau_2} f_1 (a)$
and
$f_2 ( a + m + in \tau_2 ) = e^{ \frac{\pi l}{\tau_2} m (2 a + m ) } f_2 (a)$,
respectively.
Thus, both of them do not satisfy the double periodicity condition
$f(a + m + in\tau_2 ) = f(a)$ to be satisfied by holomorphic functions
on the torus in general.
Notice, however, that for any $l \in \Z$ the former relation becomes
$f_1 ( a + m ) = f_1 ( a) $ (with $n=0$) while the
latter becomes $f_2 ( a + i n \tau_2 ) = f_2 (a)$ (with $m=0$).
This implies that each of $f_1 (a) $ and $f_2 (a)$ forms
a subset of holomorphic functions $f(a)$ that
satisfies the double periodicity condition.

We can then naturally expect the double periodicity for the quantum
holomorphic wave function $f(a)$. Indeed,
as shown in \cite{Abe:2008wn,Abe:2010sa},
we can argue that such an expectation is true for $l \in 2 \Z $.
To be more specific, a similar gauge invariance condition
on a zero-mode wave function with certain choices of
$\A^{(\tau_2 )}$ and $K$ leads to the relation
$f(a + m + i n \tau_2 ) = e^{i \pi l mn } f(a)$.
We do not make any reviews on this relation here since
we shall not utilize it in the rest of this paper.

The holomorphic wave functions $f_1 (a)$ and $f_2 (a)$
are distinct to each other but this does not mean that
they represent distinct physical states to each other. As a matter of fact,
quantum theoretically, these are both representing the same
physical state and reflect the ambiguities in the choice
of the K\"ahler potential.
This is a quintessential feature in geometric quantization.
{\it What characterizes the holomorphic zero-mode wave function
$f(a)$ is the operative relation
\beq
    \frac{\d}{\d a } f (a) \, = \, \frac{\pi l}{ \tau_2} \ba \, f(a)
    \label{3-32}
\eeq
and the inner product of the zero-mode wave functions}
\beq
    \< \Psi | \Psi^\prime \> \, = \,
    \int d\mu(a,\ba) \, e^{-K(a,\ba)} \, \overline{f(a)} f^\prime (a) \, .
    \label{3-33}
\eeq
The operative relation (\ref{3-32}) is guaranteed as long as
the K\"ahler potential is given in the form of
$K(a, \ba) = K_0 + u(a) + v (\ba )$ where $K_0 = \frac{\pi l}{\tau_2 } a \ba$.
In the present paper we have always dealt with such potentials;
see (\ref{2-22}), (\ref{3-30}) and (\ref{3-31}).
The inner product is {\it conjugate symmetric}, that is, we have
$\< \Psi | \Psi^\prime \> = \overline{\< \Psi^\prime | \Psi \>}$.
This means that there is a freedom to choose either
$f(a)$ or $\overline{f(a)}$ as the ``holomorphic'' zero-mode wave function.
Namely, at a level of definition we have a dual relation
between $f(a)$ and $\overline{f(a)}$, {\it i.e.},
\beq
    f ( a ) \, \leftrightarrow \, \overline{ f (a) } \, .
    \label{3-34}
\eeq
From this relation we can also understand the definition of $\overline{f_2 (a)}$
in (\ref{3-22}).

\section{Gauge transformations induced by modular transformations}

In the previous section we consider gauge invariance of
the zero-mode wave function in abelian CS theory on the torus
where the gauge transformations of the symplectic potential
are induced by the doubly periodic translations
$a \rightarrow a + m + i n \tau_2$ of the zero-mode variable.
In this section we continue to carry out the same analysis
for modular transformations of the zero-mode variable.
The modular transformations are generated by
combinations of the so-called modular $S$- and $T$-transformations
\beq
    S: \,  a \, \rightarrow \, - \frac{1}{a} \, , ~~~~~~ T: \,
    a \, \rightarrow \, a + 1 \,  .
    \label{4-1}
\eeq
Basics of modular forms and the modular transformations are reviewed in the next section.
Notice that the modular $T$-transformation
is obtained from the doubly periodic translations
$a \rightarrow a + m + i n \tau_2$ simply by setting $(m, n) = (1,0)$.
From (\ref{3-6}) we then find that the holomorphic zero-mode wave function $f(a)$
is invariant under the $T$-transformation regardless
the choice of the level number $l \in \Z$.
It is therefore an intriguing question whether one can
derive any conditions for $f(a)$ from the gauge invariance of
the zero-mode wave function under the gauge transformations induced
by the modular $S$-transformation.

As mentioned in the introduction, the modular $S$-transformation is not
{\it globally} well-defined since the transformation does not
preserve the double periodicity condition for the zero-mode
variable $a$ on the torus. Our analysis is based on the gauge transformations
of the symplectic potential induced by the $S$-transformation.
Thus, what we consider in this section makes sense only at {\it local} level
in terms of the zero-mode variable $a$.

We now {\it assume} that the parameter $\tau_2$ varies as
\beq
    \tau_2 \, \rightarrow \, \tau_2^\prime \, := \, \frac{\tau_2}{ \al}
    \label{4-2}
\eeq
under the modular $S$-transformation of $a$
where $\al \, ( \ne 0)$ is some constant or a function of $(a , \ba)$.
The modular transformations of our interest are thus described by
\beqar
    S: \, ( a , \tau_2 ) & \rightarrow & \left( - \frac{1}{ a} ,  \frac{\tau_2}{\al}
    \right) ,
    \nonumber \\
    T: \, ( a , \tau_2 ) & \rightarrow & (a + 1  ,  \tau_2 ) \, .
    \nonumber
\eeqar
Notice that these are qualitatively different from
the conventional $S$- and $T$-transformations used in string theory and
conformal field theory,
$S: ( \tau , z ) \rightarrow ( - \frac{1}{\tau} , \frac{z}{ \tau} )$,
$T:  ( \tau , z ) \rightarrow ( \tau + 1 , z )$ where
$\tau$ and $z$ are the modular parameter and the physical variable on
the torus, respectively.
{\it In the present case we do not focus on functions of $\tau$.
Instead, we are interested in the modular
transformations of the zero-mode complex coordinate $a$, while
$\frac{\pi l}{\tau_2}$ serves as the Planck constant $\hbar$ in the zero-mode dynamics.}

To carry out our analyses, we choose the symplectic potential $\A^{(\tau_2 ) }$
in (\ref{2-21}) and the K\"ahler potential $K_0$ in (\ref{3-3}), {\it i.e.},
\beqar
    \A^{(\tau_2 ) } &=& i \frac{ \pi l}{ \tau_2} \ba d a \, ,
    \label{4-3} \\
    K_0 &=&   \frac{\pi l}{ \tau_2} a \ba
    \label{4-4}
\eeqar
and consider the gauge invariance of the zero-mode wave function
\beq
    \Psi_0 [ a ,\ba ] \, = \, e^{- \frac{K_0}{2} } f(a) \, .
    \label{4-5}
\eeq
The gauge transformations of $\A^{(\tau_2 ) }$ are induced
by the $S$-transformation $a \rightarrow - \frac{1}{a}$ with (\ref{4-2}).
Under the $S$-transformation we have
\beqar
    K_0^\prime &  = & \frac{\al}{ |a|^4} K_0 \, ,
    \label{4-6} \\
    \del \A^{(\tau_2 )} &=& - \frac{i \pi l}{\tau_2} \left(
    \frac{ \al}{ | a|^2 } + |a|^2 \right) \frac{ d a}{a}  \, := \, d \La \, .
    \label{4-7}
\eeqar
In terms of $\La$ the gauge transformations of $\Psi_0 [ a ,\ba ]$ in (\ref{4-5})
is given by $\Psi_{0}^{\prime} \rightarrow e^{ i \La } \Psi_{0}$.
Thus the gauge invariance of $\Psi_0$ is implemented by
\beq
    e^{ i \La } e^{- \frac{K_0}{2} } f(a) \, = \, e^{- \frac{K_0^\prime}{2} }
    f \left( - \frac{1}{a} \right) \, .
    \label{4-8}
\eeq

\vskip 0.5cm \noindent
\underline{Gauge invariance of $K_0$}

We now impose the gauge invariance of $K_0$ in (\ref{4-6}), {\it i.e.},
\beq
    \al \, = \, | a |^4 \, .
    \label{4-9}
\eeq
Then $\La$ becomes $\La = -i2 K_0 \log a$ and the gauge
invariance condition (\ref{4-8}) simplifies as
\beq
    f \left( - \frac{1}{a} \right) \, = \, e^{ K_0 \log a^2 }  f(a) \, .
    \label{4-10}
\eeq
Imposition of the invariance for $K_0$ in (\ref{4-9})
is an appropriate physical condition.
Since the quantization scheme begins with the choice of $\Om^{( \tau_2 )}$,
the K\"ahler form $\Om^{( \tau_2 )}$
should physically be preserved under the  transformations of $(a , \ba)$.
In the case of the doubly periodic translations or
the modular $T$-transformations, $K_0$ is not invariant
but its variation can be split into purely holomoprhic or antiholomorphic
functions. Thus the corresponding K\"ahler form $\Om^{( \tau_2 )}$ preserves.
In the present case the transformations are simply given by $a \rightarrow - \frac{1}{a}$,
which means that the mixture of holomorphic and antiholomoprhic
parts does not come in under the $S$-transformations.
Thus the preservation of $\Om^{( \tau_2 )}$ can be
encoded by the invariance of $K_0$.

\vskip 0.5cm \noindent
\underline{Fixing the modular parameter $\tau_2$ or imposing an additional condition $|a|^2 = 1$}

In (\ref{4-2}) we have assumed that $\tau_2$ transforms under
the modular $S$-transformation of $a$. However, in principle,
it is more natural to fix the modular parameter $\tau = i \tau_2$ of the torus.
Namely, we impose $\al = 1$ or
\beq
    | a |^2 \, = \, 1
    \label{4-10a}
\eeq
such that the $S$-transformation of interest is given by
$S: ( a , \tau_2 ) \rightarrow ( -1/a , \tau_2 )$.
With the fixation of $\tau_2$ we can express the
K\"ahler potential $K_0$ as
\beq
    K_0 \, = \, \hbar |a|^2 \, = \, 1
    \label{4-11a}
\eeq
where the Planck constant (or a quantum parameter) is defined as
\beq
    \hbar \, = \, \frac{\pi l}{\tau_2 } \, := \, 1.
    \label{4-11b}
\eeq
In the unit of $\hbar = 1$ the condition (\ref{4-11a}) is equivalent to
(\ref{4-10a}).
{\it In other words, if we realize the invariance
of both $\tau_2$ and $K_0$ under the $S$-transformation of $a$, the
zero-mode variable $a$ needs to be defined
on the unit circle in the complex plane.
Under such conditions the gauge invariant condition (\ref{4-10}) can be reduced to
\beq
    \left.
    f \left( -\frac{1}{a} \right) \, = \, a^2 f(a)
    \right|_{|a|^2 = 1}
    \label{4-11}
\eeq
in the $a$-space representation. }

\section{Further analysis on the holomorphic zero-mode wave function}

\begin{figure} [htbp]
\begin{center}
\includegraphics[width=6cm]{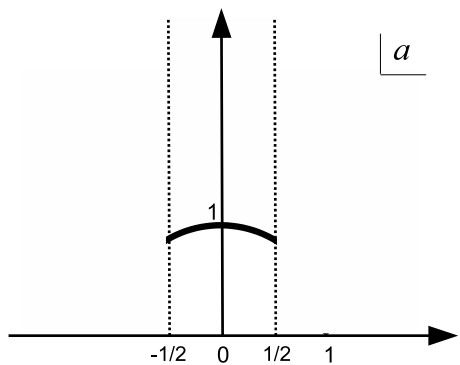}
\caption{The arc part of the boundaries in the fundamental domain ${\cal F}$, {\it i.e.}, the curve in bold
defined by $\{ a \in \C \, | ~  \im a > 0, | \re a | \le \hf , |a|^2 = 1 \}$ }
\label{fignum02v2}
\end{center}
\end{figure}

First of all we recapitulate main results in the previous section.
We consider the gauge invariance of
$f(a)$ under the gauge transformations induced by the transformations
that are relevant to the modular $S$- and $T$-transformations:
\beqar
    S_{|a|^2 = 1}: \, (a , \tau_2 ) & \rightarrow &
    \left. \left( - \frac{1}{a} , \tau_2 \right) \right|_{|a|^2 = 1} \, ,
    \label{5-10} \\
    T: \, (a , \tau_2 )  & \rightarrow & ( a  + 1 , \tau_2 ) \, .
    \label{5-11}
\eeqar
Under these transformations, the gauge invariance conditions for $f(a)$ are respectively
expressed as
\beqar
    S_{|a|^2 = 1} &:& \, \left. f \left( - \frac{1}{a} \right) =  a^2 f(a)
    \right|_{|a|^2 = 1} \, ,
    \label{5-12} \\
    T &:& \, f ( a  + 1 ) = f(a) \, .
    \label{5-13}
\eeqar
The expression (\ref{5-12}) implies that the holomorphic zero-mode wave function
$f(a)$ in abelian Chern-Simons theory on the torus behaves as a modular form of weight 2
{\it under the condition of $|a|^2 = 1$ and in the unit of
$\hbar = \frac{\pi l}{\tau_2 } = 1$}.
This means that if we were to interpret $f(a)$ as a modular form
the zero-mode variable $a$ would be defined along the arc part of the boundaries
in the fundamental domain ${\cal F}$, {\it i.e.}, $|a|^2 = 1$
with $| \re a | \le \hf$ and $\im a > 0$ (see Fig.\ref{fignum02v2});
note that basics of modular forms and the fundamental domain ${\cal F}$
(Fig.\ref{fignum02v1}) are reviewed in the appendix.
Notice also that $f(a)$ is quantum theoretically characterized by the operative relation
(\ref{3-32}) and the inner product (\ref{3-33}).

If the interpretation of $f(a)$ as a modular form of weight two is ever possible,
then what conditions do we need to impose on $f(a)$ in addition to
$|a|^2 = 1$ and $\hbar = \frac{\pi l}{\tau_2 } = 1$?
This is exactly what we are going to investigate in the present section.
To make the argument concrete, we consider the following two specific forms of $f(a)$:
\beqar
    f_1 (a ; \tau_2 ) & = & \vartheta_3 ( i \tau_2 , a ) \, = \,
    \sum_{n \in \Z} e^{- \pi \tau_2 n^2} e^{i 2 \pi n a} \, ,
    \label{5-14} \\
    f_2 (a ; \tau_2 ) & = & e^{\frac{\pi}{\tau_2}a^2} \vartheta_3 ( i \tau_2 , a )
    \nonumber \\
    & = &  \frac{1}{\sqrt{\tau_2}}
    \vartheta_3 \left( \frac{-1}{i \tau_2} , \frac{a}{i \tau_2}
    \right)
    \, = \,
    \frac{1}{\sqrt{\tau_2}} \sum_{n \in \Z} e^{- \frac{\pi}{\tau_2} n^2 + \frac{ 2 \pi}{\tau_2} n a}
    \label{5-15}
\eeqar
where in (\ref{5-15}) we use the formula (with $\re \, \al > 0$)
\beq
    \sum_{m \in \Z} e^{-\pi \al m^2 + 2 \pi i \bt m} \, = \,
    \frac{1}{\sqrt{\al}} \sum_{n \in \Z} e^{- \frac{\pi}{\al}( n - \bt )^2} \, .
    \label{5-15a}
\eeq
Both $f_1 (a)$ and $f_2 (a)$ are the same as those defined in section 3.

\vskip 0.5cm \noindent
\underline{The modular $T$-invariance}

From the relation (\ref{3-9}) it is obvious that
$f_1 (a)$ satisfies the modular $T$-invariance (\ref{5-13}).
On the other hand, as mentioned below (\ref{3-31}),
$f_2 (a)$ does not satisfy $f(a + 1) = f(a)$ but $f (a + i \tau_2 ) = f(a)$.
Indeed, we find $f_2 (a+1 ) =
    e^{\frac{\pi}{\tau_2}(a + 1)^2} \vartheta_3 ( i \tau_2 , a + 1 )
      =   e^{\frac{\pi}{\tau_2}(2a + 1)} f_2 ( a )$
and
$f_2 (a + i \tau_2 ) = \frac{1}{\sqrt{\tau_2}}
    \vartheta_3 \left( \frac{-1}{i \tau_2} , \frac{a + i \tau_2 }{i \tau_2}
    \right) = \frac{1}{\sqrt{\tau_2}}
    \vartheta_3 \left( \frac{-1}{i \tau_2} , \frac{a}{i \tau_2}
    \right) = f_2 (a)$.
Thus the modular $T$-invariance (\ref{5-13}) is not realized by $f_2 (a)$, {\it per se}.
However, as discussed earlier, $f_1 (a)$ and $f_2 (a)$ are related to each other
up to the purely holomorphic (or anti-holomorphic) choice of the K\"{a}hler potential.
In this sense, we find that the holomorphic zero-mode wave function $f(a)$ satisfies the
modular $T$-invariance (\ref{5-13}).

\vskip 0.5cm \noindent
\underline{The modular $S_{|a|^2 = 1}$-invariance}

Contrary to the $T$-invariance (\ref{5-13}), it is not straightforward to see the
the $S_{|a|^2 = 1}$-invariance (\ref{5-12}) for either of $f_i (a)$ $(i=1,2)$.
From the specific form (\ref{5-15}, \ref{5-15a}) of $f_i (a)$ we find
\beq
    \overline{f_i (a ) } = f_i \left( \frac{1}{a} \right)
    = f_i \left( \frac{1}{a} \right)
    \label{5-16}
\eeq
for $|a|^2 = 1$.
Bearing in mind that the holomorphic wave function $f(a)$ is
characterized by the operative relation (\ref{3-32}), we then
find that the $S_{|a|^2 = 1}$-invariance (\ref{5-12}) can be
realized by imposing the following condition on $f(a)$:
\beq
    \d \overline{f(a)} = \bd f(a)
    \label{5-17}
\eeq
where $\bd = \frac{\d}{\d \ba} = \d_\ba$.
Notice that from the operative relation (\ref{3-32}) with $\hbar = \frac{\pi l}{\tau_2 } = 1$
this can be expressed as $ \ba \overline{f(a)} = a f(a)$; thus (\ref{5-16}) straightforwardly
leads to the $S_{|a|^2 = 1}$-invariance (\ref{5-12}).
Note also that we can directly check the operative relation $\bd f(a) = a f(a)$
in the case of $f_1 (a)$ under $|a|^2 = 1$ as follows.
\beqar
 \frac{\d}{\d \ba} f_1 \left(  a \right)
 &=& \frac{\d}{\d \ba} f_1 \left(  \frac{1}{\ba} \right)
 \nonumber \\
 &=&
 \d_\ba  \sum_{n \in \Z} e^{- \pi \tau_2 n^2} e^{i 2 \pi n \frac{1}{\ba} }
 ~ = ~
 \sum_{n \in \Z} \frac{i 2 \pi n }{\ba^2} e^{- \pi \tau_2 n^2} e^{i 2 \pi n \frac{1}{\ba} }
 \nonumber \\
 &=&
 \frac{1}{\ba^2} \d_a f_1 (a)
 ~ = ~
 \frac{1}{\ba^2} \ba f_1 (a)  ~ = ~ a f_1 (a)
 \label{5-18}
\eeqar

From an ordinary classical point of view the imposing condition (\ref{5-17})
means that $\bd f(a)$ is real but quantum theoretically
$\bd f(a) = a f(a)$ ($a \in \C$) is not necessarily real in general.
Quantum theoretically, the condition (\ref{5-17}) is in accord with
the the conjugate symmetry of the inner product and
the consequent dual relation of the holomorphic wave function
$f(a) \leftrightarrow \overline{f(a)}$ as discussed in (\ref{3-34}).

\section{Conclusion}

In this paper we consider a wave function in abelian Chern-Simons
theory on the torus in the context of geometric quantization.
In a suitable gauge the wave function can be parametrized by
a zero-mode variable $a$ on the torus, with a modular parameter of the torus
chosen by $\tau = i \tau_2$ ($\tau_2 > 0$).
All the key ingredients in the geometric quantization,
such as a K\"ahler potential $K(a ,\ba )$, a symplectic potential
$\A^{( \tau_2 )}$ and a zero-mode wave function $\Psi[ a , \ba ]$,
can all be derived from the K\"ahler form $\Om^{( \tau_2 )}$
for the zero-mode variable.
In this sense the geometric quantization of our interest can be
considered as a K\"ahler-form program.
In general, the wave function satisfies a polarization condition.
This allows us to express it as
$\Psi [ a , \ba ] = e^{-\frac{K(a ,\ba )}{2}} f(a)$
where $f(a)$ is a holomorphic function which we call
a holomorphic zero-mode wave function.
In this paper we emphasize that there exist ambiguities
in the choice of $K ( a , \ba )$ up to an addition of holomorphic
and antiholomorphic functions. We review these materials in section 2.

In section 3 we consider the gauge invariance of $\Psi [ a , \ba ]$, where
the gauge transformations of $\A^{(\tau_2 )}$ are induced by the doubly periodic
translations $a \rightarrow a + m + i n \tau_2$, and find that
$f(a)$ can be described in terms of the Jacobi theta functions.
Along the way, we indicate that
$f(a)$ is quantum theoretically determined by the operative relation (\ref{3-32})
and the inner product of the zero-mode wave functions (\ref{3-33}).

In section 4 we make a similar argument on the gauge invariance
of $\Psi [ a , \ba ]$ when the gauge transformations of
$\A^{(\tau_2 )}$ are induced by the modular transformations of $a$.
Under the condition of $|a|^2 = 1$ and
in the unit of $\hbar = \frac{\pi l}{\tau_2} = 1$
we find that $f(a)$ behaves as a modular form of weight 2.

In section 5 we further study how exactly one can interpret
$f(a)$ as the modular form of weight 2 by use of specific
representations of $f(a)$ in terms of the Jacobi theta functions.
We find that in order for such an interpretation to be ever possible,
we need to impose an additional condition
$\d \overline{f(a)} = \bd f(a)$ on $f(a)$ besides
$|a|^2 = 1$ and $\frac{\pi l}{\tau_2} = 1$.
We have pointed out that his additional condition is
in accord with the dual relation of the holomorphic wave function
$f(a) \leftrightarrow \overline{f(a)}$ buried in the definition
of the inner product (\ref{3-33}) but we can not find any
satisfactory reasonings for (\ref{5-17}), except that it leads to
(\ref{5-12}), in the present paper.
Also, the crucial condition $|a|^2 = 1$ indicates that
the interpretation of the modular form only makes sense
along the arc part of the fundamental domain, as shown in Figure \ref{fignum02v1}.
Therefore we can not fully relate $f(a)$ to the modular form of weight 2.
In the present paper, however, we show that such a relation
is partly possible under certain conditions on $f(a)$ and $a$.

\vskip .3in
\noindent
{\bf Acknowledgments} \vskip .04in \noindent
The author would like to thank anonymous referees of a certain journal
to which a previous version of this preprint (arXiv:1711.07122v2 [hep-th]) is submitted.
The refereeing reports help the author improve the original manuscript very much.

\appendix
\section{Basics of modular forms}

The last expression (\ref{4-11}) in section 4
implies that under certain conditions, in particular with $|a| = 1$, the
holomorphic zero-mode wave function $f(a)$ in abelian Chern-Simons (CS) theory
on the torus behaves as a modular form of weight 2.
In order to further study this intriguing result, in this appendix
we briefly review some basics of the modular forms, following mathematical
textbooks \cite{Koblitz:1993bk,Kurokawa:2005bk,Lozano-Robledo:2011bk}.

\begin{figure} [htbp]
\begin{center}
\includegraphics[width=7cm]{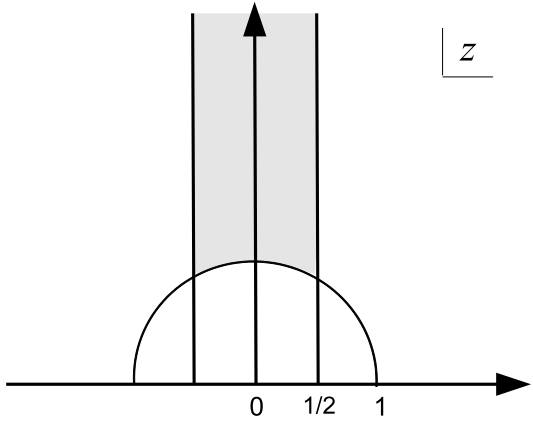}
\caption{Fundamental domain ${\cal F} $ for the action of the $SL (2 , \Z)$ generator
 on the upper-half plane $\H$}
\label{fignum02v1}
\end{center}
\end{figure}

In general, a modular form $f(z)$ of weight $k$ is defined by
\beq
    f \left( \frac{ \al z + \bt }{\ga z + \del } \right) \, = \, ( \ga z + \del )^k  f(z)
    \label{5-1}
\eeq
where $\al$, $\bt$, $\ga$, $\del$ are matrix elements of the modular group
\beq
    SL( 2, \Z ) \, = \, \left\{ \left.
    \left(
      \begin{array}{cc}
        \al & \bt \\
        \ga & \del \\
      \end{array}
    \right) \right| \, \al, \bt, \ga, \del \in \Z , ~ \al \del - \bt \ga = 1
    \right\} \, := \, \Ga  \, .
    \label{5-2}
\eeq
The modular forms are defined on the upper-half plane
$\H = \{ z \in \C \, | \, \im \, z > 0 \}$.
Accordingly, to be rigorous, the modular group can be defined as
$PSL( 2 , \Z ) := SL( 2 , \Z ) / \{ \pm I \}$, with $I$ the identity matrix.
The fundamental domain ${\cal F}$ for
the action of $SL (2 , \Z)$ generators on $\H$ corresponds to
the space of $\H / PSL( 2 , \Z ) $.
This can be specified by
\beq
    {\cal F} \, = \, \left\{ z \in \C \left| \,  \im \, z > 0 , ~
    |z| \ge 1, ~ |\re \, z | \le \hf \right. \right\} \, .
    \label{5-3}
\eeq
This region is shown in Figure \ref{fignum02v1}.
Any point $z \in \H$ can be obtained by a linear fractional transformation
of $z_0 \in {\cal F}$, {\it i.e.}, $z = \frac{\al z_0 + \bt }{ \ga z_0 + \del}$.
In this sense we can properly consider $\Ga  = SL ( 2 , \Z )$ in (\ref{5-2}) as the
modular group.

It is well known that the modular group can be generated by
$\left( \!
  \begin{array}{cc}
    1 & 1 \\
    0 & 1 \\
  \end{array}
\! \right)$ and $
\left( \!
  \begin{array}{cc}
    0 & -1 \\
    1 & 0 \\
  \end{array}
\! \right)$.
The definition of the modular form in (\ref{5-1}) is then obtained
from the conditions
\beqar
    f ( z + 1 ) &=& f(z) \, ,
    \label{5-4} \\
    f \left( - \frac{1}{z} \right) &=& z^k f( z) \, .
    \label{5-5}
\eeqar
The first condition simply means that
$f(z)$ can be expressed in a form of the Fourier expansion
\beq
    f(z) \, = \, \sum_{n = 0}^{\infty} a_n \, q^{n}
    \label{5-6}
\eeq
where $q = e^{i 2 \pi z }$ and $a_n$ is the Fourier coefficient.
If $a_0 = 0$, the modular form $f(z)$ is called the {\it cusp form}.
The vector space formed by the cusp forms of weight $k$ is denoted
by $S_k ( \Ga )$, {\it i.e.},
\beq
    S_k ( \Ga ) \, := \, \left\{
    f: \H \rightarrow \C \left| \,
    f \left( - \frac{1}{z} \right) = z^k f( z) , \,
    f(z) =  \sum_{n = 1}^{\infty} a_n \, q^{n}
    \right.
    \right\} .
    \label{5-7}
\eeq

Let $f(z) , \, g(z) \in S_k (\Ga )$, then the so-called
{\it Petersson inner product} is defined as
\beq
    \< f , g \> \, = \, \frac{1}{{\rm vol}  {\cal F}  } \int_{\cal F}
    f(z) \overline{ g (z) } \, y^k \, \frac{dx  dy }{y^2}
    \label{5-8}
\eeq
where $z = x + i y$.
Notice that the integral measure $\frac{dx dy}{y^2}$
is invariant under the modular transformations.
Invariance under the $T$-transformation is obvious.
Under the $S$-transformation
we have $y^\prime = \frac{y}{ | z |^2}$ and
the unit area is changes as $d x^\prime dy^\prime = \frac{dx dy}{|z|^4}$.
Thus the measure $\frac{dx dy}{y^2}$ is modular invariant.
This measure is called the {\it hyperbolic measure} on $\H$.

The volume of ${\cal F}$ is evaluated with the hyperbolic measure and
becomes finite:
\beq
    {\rm vol}{\cal F} \, = \, \int_{\cal F} \frac{dx  dy }{y^2}
    \, = \, \int_{-\hf}^{\hf} \left( \int_{\sqrt{1-x^2}}^{\infty} \frac{dy}{y^2} \right) dx
    \, = \, \int_{-\hf}^{\hf} \frac{dx}{\sqrt{1- x^2} } \, = \, \frac{\pi }{3} \, .
    \label{5-9}
\eeq
Notice also that the integrand of the Petersson inner product,
$f (z) \overline{g(z) } y^k$, is invariant under both the $S$- and $T$-transformations.
Hence, the inner product (\ref{5-8}) represents
a manifestly modular invariant integral.

\vskip .3in

\end{document}